\documentstyle[epsfig]{article}

\def\frac#1#2{ {{#1} \over {#2} }}

\def\half{\mbox{\small $\frac{1}{2}$}}

\def\eg{\hbox{\rm e.g. }}

\def\abs#1{\left| \: #1 \: \right|}%
\def\bom#1{\mbox{\boldmath$#1$}}
\def\beq{\begin{displaymath}}
\def\eeq{\end{displaymath}}
\def\re#1{(\ref{#1})}
\def\ee{$e^+e^-\;$}
\def\as{\alpha_S}
\def\asb{\bar \alpha_S}

\def\bk{\bom {k} }

\def\bkq{\abs{\bom{k}+\bom{q}}}
\def\om{\omega}
\def\ga{\gamma}
\def\tga{\tilde \gamma}
\def\tchi{\tilde \chi}
\def\tom{\tilde \om}
\def\de{\delta}
\def\De{\Delta}
\def\cF{{\cal F}}
\def\cA{{\cal A}}
\def\Q_s{\mu}
\def\np#1#2#3{Nucl.\ Phys.\ {\bf B#1}, #2 (19#3)}
\def\pl#1#2#3{Phys.\ Lett.\ {\bf #1B}, #2 (19#3)}
\def\pr#1#2#3{Phys.\ Rev.\ D {\bf #1}, #2 (19#3)}
\def\prep#1#2#3{Phys.\ Rep.\ {\bf #1}, #2 (19#3)}

\def\zp#1#2#3{Zeit.\ Phys.\ {\bf C#1}, #2 (19#3)}
\title{Angular ordering and structure functions at small
  $x$\footnote{Talk given at the ``'', Trieste, 12-16 May 1997}}

\author{G. Bottazzi \address{ Dipartimento di Fisica Universit\`a di
    Milano and INFN Sezione di Milano\\ Via Celoria 16, 20133 Milano,
    Italy}
}
       
\begin{document}
\begin{flushright}
IFUM 575-FT \\
hep-ph/9708474\\[12pt]
\end{flushright}

\begin{center}
  \parskip=0pt
  {\LARGE \bf Angular ordering and structure functions at small $x$
\footnote{Talk given at the 5th Topical Seminar on ``The Irresistible
  Rise of the Standard Model'', San Miniato, Italy, 21-25 April 1997}}

  \bigskip

  {\large  G. Bottazzi}

  \bigskip  

  Dipartimento di Fisica Universit\`a di
  Milano and INFN Sezione di Milano\\ Via Celoria 16, 20133 Milano
  Italy
\end{center}

{\small
We examine the effect of using an angular ordered evolution equation
for the unintegrated gluon distribution at small $x$.
}


\section{Introduction}
Angular ordering has proven to be a universal aspect of high energy
processes. The physical origin of this ordering \cite{IR} is the
coherence of the soft emissions: namely the cancellation outside the
ordered region of multiple soft gluons emission.  The universality
consists in the fact that this cancellation works both in time-like
processes (\eg \ee annihilation) and in space-like processes (\eg deep
inelastic scattering).
Due to its universal validity this description can in principle
constitute a bridge between the usual finite-energy (DGLAP) and the
soft exchange (BFKL) description \cite{BFKL}.

The detailed analysis of angular ordering in multi-parton emission at
small Bjorken-$x$ and the related virtual corrections has been done in
Ref.~\cite{CCFM} (see also \cite{March}), where it was shown that to
leading order the initial-state gluon radiation can be formulated as a
branching process in which angular ordering in both real emissions and
virtual corrections is taken into account.

In the totally inclusive sum which defines the gluon density, the
higher collinear singularities which come both from real and virtual
contributions cancel.  As a result, to leading order the small-$x$
gluon density is obtained by resumming $\ln x$ powers coming only from
IR singularities, and angular ordering contributes only to subleading
corrections.

The calculation of the gluon density by resummation of $\ln x$ powers
without angular ordering was done $20$ years ago \cite{BFKL} and led
to the BFKL equation.  The solution of this equation, $\cF(x,k)$, is
the unintegrated gluon density at fixed transverse momentum $k$
and is related to the small-$x$ part of the gluon structure function
$F(x,Q)$ by
\begin{equation}\label{cF}
F(x,Q)=\int d^2 \bk \;\cF(x,k)\theta(Q-k)\,.
\end{equation}
In this talk,
as a first step of a systematic study of multi-parton emission in DIS, 
the effect of angular ordering on the small-$x$ evolution of the gluon 
structure function is studied, with both analytical and numerical
techniques \cite{bmss}. 

\section{Evolution equation for the gluon density}

In this section we recall the basic ingredients used to build the
coherent branching equation for the gluon density at small $x$.  The
evolution of the gluon density is thought as a multi-branching process
involving only gluons, since gluons dominate the small-$x$ region
(fermion exchange is subleading by a factor $\log x$).
Consider the single branching process pictorially represented in
fig.~1.  We denote by $x_i$ and $\bk_i$ respectively the energy
fraction and transverse momentum of the $i$-th exchanged
(t-channel) gluon. The energy fraction of the $i$-th emitted
gluon is $(1-z_i)x_{i-1}$, where $z_i=x_i/x_{i-1}$, and
$\bom q_i$ denotes its transverse momentum. In what follows we shall
use the notation $k'=\bkq$. With each branching is associated
the factor:
\begin{equation}\label{dP}
d{\cal P}_i
=\frac{d^2q_i}{\pi q_i^2} \,\frac{\asb dz_i}{z_i}
\,\De(z_i,q_i,k_i)\,\theta(\frac{q_i}{z_{i-1}}-q_{i-1})
\end{equation}
where the function $\De$
\begin{equation}\label{De}
\ln \De(z_i,q_i,k_i)=
-\int_{z_i}^1 \,\frac{\asb \, dz'}{z'}
\int\frac{dq'^2}{q'^2}\,\theta(k_i-q')\,\theta(q'-z'q_i)
\end{equation}
is the form factor which resums IR singularities (small $z_i$) which
appear in virtual corrections to soft exchanged gluons.  This form
factor corresponds, in the BFKL equation, to the gluon Regge form
factor.
\begin{figure}[t]\label{gpsfig:kinebw}
\begin{center}
%under unix
    \scalebox{1.0}{\input{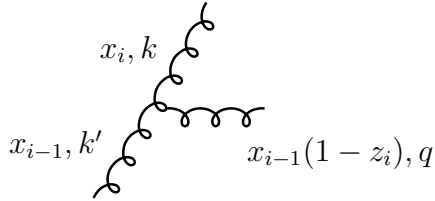}}
%under dos
%    \scalebox{1.0}{\input{branch.pst}}
    \caption[]{
      $k_i$ are the gluons exchanged along the t-channel and $q_i$ the
      emitted ones.  $k_{i}'=\abs{\bom{k_i}+\bom{q_i}}$}
\end{center}
\end{figure}

To see where the angular ordering is in the splitting
kernel $d{\cal P}$ and in the form factor $\De$ we can write down the
condition $\theta_{i}>\theta_{i-1}$, where $\theta_{i}$ is the angle
of the $i$-th emitted gluon with respect to the initial incoming gluon.
In terms of the transverse momenta, the constraint will read
$ q_{i} > z_{i-1} q_{i-1}$ and is exactly what we find in \re{dP} and
\re{De}.  The so defined branching process is accurate to leading IR
order and, at the inclusive level, does not require any collinear
approximation (for a proof see \cite{CCFM}).

Angular ordering provides a lower bound on transverse momenta so
that no collinear cutoff is needed other than virtuality for
the first incoming gluon. On the other hand, in order to deduce a
recurrence relation for the inclusive distribution, one has to
introduce a ``maximum available angle'' $\bar\theta$ in the form of a
transverse momentum $p$.  If $x_n$ and $k_n$ are the kinematical
variables of the last exchanged gluon (nearest to the hard vertex) the
constraint $ \theta_n <\bar \theta $ becomes $z_nq_n < p \simeq
x_n E\bar\theta$ where $E$ is the incoming parton energy.

The distribution for emitting $n$ initial state gluons in defined as 
\begin{equation}\label{An}
\cA^{(n)} (x,k,p) \;=\;
\int \prod_{i=1}^{n} \; d {\cal P}_i \;
\theta(p-z_nq_n) \; \de(k^2-k_n^2) \; \de(x-x_n)
\end{equation}
so that the fully inclusive gluon density becomes
\begin{equation}\label{A}
\cA(x,k,p) \;=\; \sum_{n=0}^{\infty} \; \cA^{(n)} (x,k,p)\,.
\end{equation}
This satisfies the so called CCFM equation \cite{CCFM}:
\begin{eqnarray}\label{A1}
&\cA(x,k,p) \,=\, \cA^{(0)}(x,k,p) \,+\, \\
&\int\frac{d^2q}{\pi q^2}\,
\int_x^1\frac{\asb\,dz}{z^2}\,\De(z,q,k)\,
\theta(p-zq)\;\cA(\frac{x}{z},k',q)
\nonumber
\end{eqnarray}
where the inhomogeneous term $ \cA^{(0)}(x,k,p) $ is the
distribution for no gluon emission.

This equation can be partially diagonalised by introducing the
$\om$-representation
\begin{equation}
\cA_{\om}(k,p) \;=\; \int_0^1 dx \; x^{\om}\cA(x,k,p)
\,.
\end{equation}
so that \re{A1} becomes
\begin{eqnarray}\label{A2}
&\cA_{\om}(k,p) \;=\; \cA_{\om}^{(0)}(k,p) \;+\; \\
&\int\frac{d^2q}{\pi q^2}\,\int_0^1\frac{dz\,\asb\,z^{\om}}{z}\De(z,q,k)\,
\theta(\frac{p}{z}-q)\,\cA_{\om}(k',q)\,.
\nonumber
\end{eqnarray}
No further diagonalisation in transverse momentum is possible since
the kernel depends both on the total momentum $k$ and on $q$ and $p$.

In order to explicitly see the collinear safety of 
the fully inclusive gluon density $\cA(x,k,p)$ it is possible to
rewrite the evolution equation in an ``inclusive'' form:
\begin{eqnarray}\label{A3}
&&\cA_{\om}(k,p) \,=\, \tilde \cA_{\om}^{(0)}(k,p) \,
+\, \de_{\om}(k,p) \,+ \\
&&\frac{\asb}{\om}\,\int\frac{d^2q}{\pi q^2}\,
\left[ \cA_{\om}(k',q)\,-
\theta(k-q)\,\cA_{\om}(k,\tilde{q})\right]
\nonumber
\end{eqnarray}
where the inhomogeneous term is
\begin{equation}
\label{A3b}
\tilde \cA_{\om}^{(0)}(k,p) =
\cA_{\om}^{(0)}(k,p)
+\frac{\asb}{\om}\int^k\frac{dq^2}{q^2}\cA^{(0)}_{\om}(k,\tilde{q})
\,,
\end{equation}
with $\tilde{q}= \hbox{min} (q,p)$. Forgetting the extra factor
$\de_{\om}(k,p)$, which is collinear safe, one sees explicitly that
\re{A3} is finite in the $q\to0$ collinear limit.

\subsection{The BFKL limit.}

For $p\to\infty$ the term $\de_{\om}(k,p)$ vanishes and
the gluon density $\cA_{\om}(k,p)$ becomes independent of $p$.
In this limit the equations \re{A3} becomes
\begin{eqnarray}\label{bfkl}
&\cF_{\om}(k) \;=\; \tilde \cF_{\om}^{(0)}(k) \;+\; \\
&\frac{\asb}{\om} \int\frac{d^2q}{\pi q^2}\;
\left[\cF_{\om}(k') \,-\, \theta(k-q)\,\cF_{\om}(k)\right]
\nonumber
\end{eqnarray}
which is the BFKL equation for the gluon density that possesses the well
known solution
\begin{equation}\label{bfkl2}
\cF_{\om}(k)=\int_{\frac12 - i \infty}^{\frac12 + i \infty}
\frac{d\ga}{2\pi i}
\,\frac{1}{k^2}\left(\frac{k^2}{k^2_0}\right)^{\ga}
\frac{\om f_0(\om,\ga)}{\om-\asb\chi(\ga)}
\end{equation}
where $k_0$ and the function $f_0$ are fixed by the inhomogeneous term
and $\chi(\ga)$ is the BFKL characteristic function $\chi(\ga)= 2
\psi(1)-\psi(\ga)-\psi(1-\ga)$. For a general initial condition the
asymptotic behaviour of $\cF_{\om}(k)$ for $k \gg k_0$ and for $k \ll
k_0$ is given by solving the characteristic equation
$\omega=\asb\chi(\ga)$ in the regions $0<\ga<\half$ and $\half<\ga<1$
respectively. Using the saddle point approximation around the leading
singularity of $\ga(\as/\om)$ in the $\om$-plane, which is at
$\ga_c=\ga(\as/{\om_c})=1/2$, one obtains the asymptotic behaviour of
$\cF(x,k)$ at small $x$ which reads:
\begin{equation}\label{asy}
x \cF(x,k) \sim
\;\frac{x^{-\om_c}}{k^2}\left(\frac{k^2}{k^2_0}\right)^{\ga_c}
\,,
\end{equation}
where $\om_c=\asb\chi(\half)=4\asb\ln2$.

\subsection{Properties of the gluon distribution.}

Consider a solution to \re{A2} of the form:
\begin{equation}\label{cAom}
\cA_{\om}(k,p) =
\;\frac{1}{k^2}\left(\frac{k^2}{k^2_0}\right)^{\tga}
\;G\left(\frac{p}{k}\right)
\,,
\end{equation}
with $\tga$ a function of $\as$ and $\om$.
This form allows a direct comparison with the BFKL
asymptotic behaviour (see \re{asy}). With $0<\tga<1$, from \re{A2}
one finds the equation:
\begin{eqnarray}\label{dG}
&p\;\partial_p\;G(p/k)= \\
&\asb
\int_p\,\frac{d^2q}{\pi q^2}
\,\left(\frac{p}{q}\right)^{\om} \De(\frac{p}{q},q,k)
\,G(\frac{q}{k'})
\,\left(\frac{{k'}^2}{k^2}\right)^{\tga-1}
\nonumber
\end{eqnarray}
where we have chosen the boundary condition $G(\infty)=1$. The
function $\tga$ must satisfies the ``characteristic'' equation:
\begin{eqnarray}\label{A5}
&1=\frac{\asb}{\om}\tchi(\tga,\as) &\\
&\tchi=\int\frac{d^2q}{\pi q^2}
\left\{
\left(\frac{{k'}^2}{k^2}\right)^{\tga-1}
\!\!\!G(\frac{q}{k'})-\theta(k-q)\,G(\frac{q}{k})\right\}.&
\nonumber
\end{eqnarray} 
obtained from \re{A3} in the $p \to \infty$ limit.

Notice that when $\asb \to 0$ the function $G$ reduces to a constant
and this implies that $\tchi$ becomes the BFKL
characteristic function. Since $1-G$ is of order $\asb$ and
doesn't have the $1/\omega$ enhancement factor, we can conclude that
the effect of angular ordering on the asymptotic behaviour is of subleading
nature. Moreover it's possible to show \cite{bmss} that in the limit $\ga \to 0$
the difference $\chi(\ga)-\tchi(\ga,\as)$ tends to a constant (in
$\as$) implying a behaviour of the form $\as^3/{\omega^2}$.

As we shall see from the numerical analysis, the characteristic
function $\tchi(\tga,\as)$ decreases with $\tga$, reaches a minimum at
$\tga_c<1$ (for reasonable $\as$), and then rises again.  As in the
BFKL case, we shall denote by $\tom_c$ the leading singularity in
$\om$ which corresponds to the minimum of the characteristic function
at $\tga=\tga_c$.

\section{Numerical results.}

\begin{figure}[!t]
\begin{center}
\epsfig{file=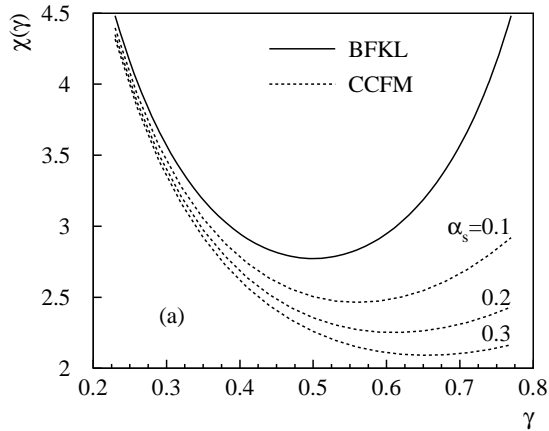,width=7.5cm}
\caption[]{The characteristic function $\tchi(\ga,\as)$ for different
  $\as$ compared to $\chi(\ga)$ plotted as functions of $\ga$}
\label{fig:chi}
\end{center}
\end{figure}
Fig.~\ref{fig:chi} shows the results for $\tchi$ as a function of
$\tga$ for various $\as$.  The difference $\de \chi=\chi-\tchi$ is
positive and increases with $\tga$ and with $\as$.  Moreover we find $\de
\chi \sim \tga$ for $\tga\to0$ ($\asb$ small and fixed) and $\de \chi
\sim \asb$ for $\asb\to0$ ($\tga$ small and fixed) as expected from
analytical studies.
\begin{figure}[!t]
\begin{center}
\epsfig{file=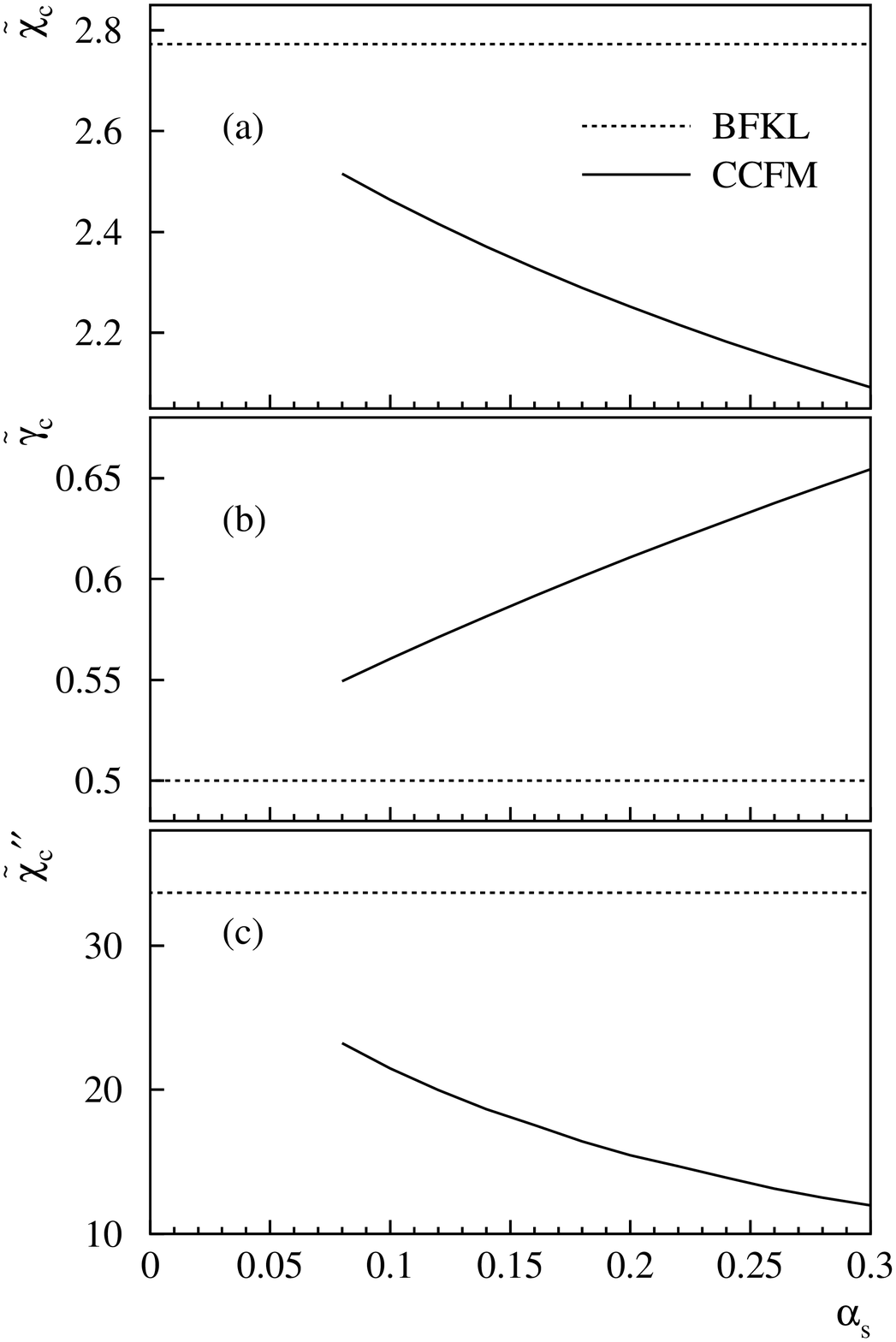,width=7.5cm}
\caption[]{(a) The value of the minimum of the characteristic function
  $\tchi_c$, as a function of $\as$. (b) The position of the minimum
  of the characteristic function $\tga_c$, as a function of $\as$.
  (c) The second derivative of the characteristic function
  ${\tchi_c}''$ at its minimum, as a function of $\as$.}
\end{center}
\end{figure}
The function $\tchi$ decreases faster than $\chi$ for increasing
$\tga$: the minimum of the characteristic function gets shifted to the
right and is lower.  This produces a milder
growth of the structure function at small $x$ and moderate $\bk$.  In
Fig.~3a and 3b we plot as a function of $\as$ the values $\tchi_c$ and
$\tga_c$ of the minimum of $\tchi$ and its position $\tga_c$.  As
expected the differences compared to the BFKL values $\chi_c=4\ln 2$
and $\ga_c=\half$ are of order $\asb$.
In Fig.~3c we show the second derivative, ${\tchi_c}''$, of the
characteristic function at its minimum.The diffusion in $\ln k$ is
inversely proportional to the square root of ${\tchi_c}''$
\footnote{This is strictly true only for the solution in the
  saddle-point approximation, nevertheless this quantity remain a good
  indicator due to the mild asymptotic behaviour of the $G$ function.}.
One can see therefore that the inclusion of angular ordering
significantly reduces the diffusion compared to the BFKL case.

The loss of symmetry under $\ga\to1-\ga$, which clearly shows up in
Fig.~2, relates to the loss of symmetry between small and large scales:
while in BFKL the regions of small and large momenta are equally
important, the angular ordering favours instead the
region of larger $k$.  However, at each intermediate branching, the
region of vanishing momentum is still reachable for $x\to0$, so that
the evolution still contains non-perturbative components.

The subleading nature of the angular ordering inclusion in the
evolution equation that we have seen so far is no longer guaranteed if
one considers, instead of the gluon density, more exclusive quantities. This
happens because the cancellations between real emissions and virtual
corrections which, in the angular ordering equation, reconstruct at
leading level the BFKL solution no longer work for a modified kernel,
such the one used for associated distributions.

The topic requires further analysis but for now let us show some
preliminary plots. In Fig.~4a and Fig.~4b we show the ``rung
multiplicity'' and the transverse momentum flow calculated for given
kinematical variables. Notice that even if for the chosen values of
$x$ and $k_t$ the gluon distribution calculated in the two approaches
are practically the same, the distribution shapes and normalisation
are quite different for the BFKL case compared to the CCFM equation.

\begin{figure*}[t]
\begin{center}
\epsfig{file=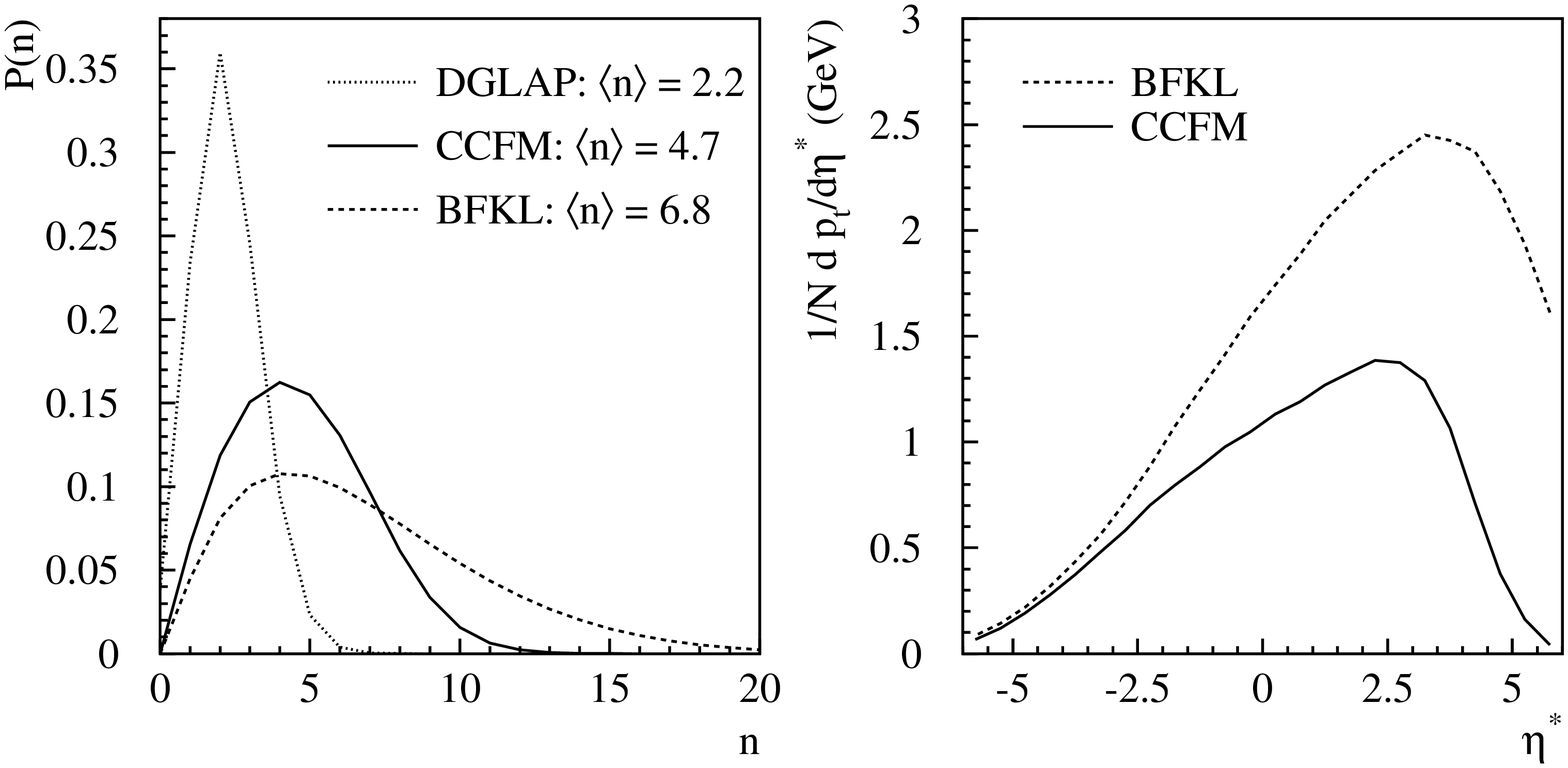,width=12cm}
\caption[]{(a) Distribution of number of emission with $q>q_0=1$GeV, for
  DGLAP, CCFM and BFKL evolution to $x = 5.10^{-5}$, $k=5\;$GeV,
  $\as=0.2$.  (b) Transverse momentum flow in the hadronic centre of
  mass frame as a function of the rapidity $\eta^*$ for evolution to
  $x=2\cdot 10^{-4}$, $k=3\;\hbox{GeV}$, $\as=0.2$ (the proton
  direction is to the left).}
\label{fig:figass}
\end{center}
\end{figure*}

\section*{Acknowledgements}

This research was carried out in collaboration with G.~Marchesini,
G.P.~Salam and M.~Scorletti and supported in part by the Italian MURST.

\end{document}